# CAWS-Security Algorithms for Wireless Sensor Networks: A Cellular Automata Based Approach

Nilanjan Sen[1] and Indrajit Banerjee[2]

*Abstract*— **Security in the Wireless Sensor Networks (WSN) is a very challenging task because of their dissimilarities with the conventional wireless networks. The related works so far have been done have tried to solve the problem keeping in the mind the constraints of WSNs. In this paper we have proposed a set of cellular automata based security algorithms (CAWS) which consists of CAKD, a Cellular Automata (CA) based key management algorithm and CASC, a CA based secure data communication algorithm, which require very small amount of memory as well as simple computation.**

## I. INTRODUCTION

THE Wireless Sensor Network (WSN) has become a very emerging field in computer science now-a-days due to its multifarious applicability that includes military applications, industrial applications, home applications, monitoring temperature, humidity, vehicular movement etc. [1]. Besides this, as there are lot of areas in it that can be improved further, it has become a very interesting field of research also.

The security threats of a WSN are different than conventional mobile networks. A large number of sensors are used here compared to other mobile networks. They can be extended or curtailed dynamically because here the sensors can be added or removed easily after the deployment. Not only these, the sensors may be deployed in the hostile areas that increase the chance of attacks by the adversaries [2]. That is why the research methodologies for the security of WSNs are little bit different also. According to the Perrig et al [3] the security properties of the sensor networks are data confidentiality, data authentication, data integrity and data freshness. Different researchers are currently doing research in the field of WSN's node management, data authentication, data security, node integrity etc. In this paper we have proposed a set of algorithms collectively known as CAWS for the key pre-distribution and secure data communication using Cellular Automata (CA).

[1] Purabi Das School of Information Technology, Bengal Engineering and Science University, Shibpur; nilanjansenin@yahoo.co.in

[2] Department of Information Technology, Bengal Engineering and Science University, Shibpur; ibanerjee@it.becs.ac.in

The remaining portion of this paper is organized as below. Section II provides a brief overview of WSN, CA and Reversible CA. In Section III, we have discussed the related research works in brief. Section IV and V contains CAKD and CASC, i.e. our proposed key pre-distribution algorithm and secured data communication algorithm respectively. In Section VI, we have analyzed our algorithms. Finally, we have concluded our paper in Section VII.

## II. PRELIMINARIES

### A. Wireless Sensor Network (WSN)

WSNs are composed of a large number of sensor nodes. Generally WSN architecture has four key features: i) it is self-organized, ii) it can take the decision locally, iii) it supports wireless communication and iv) its traffic flow is almost unidirectional towards the destination or 'sink' [4]. A sensor node has embedded processor and storage system. But unlike traditional wireless network, sensor devices in WSN have limited computation and communication capabilities and limited energy because they are battery powered. The sensor nodes can collect the data, analyze them, and send them to a destination. The domain of the sensor nodes is known as sensor field. They communicate with each other as well as with their neighboring domains normally through wireless connection. It acquires the information of its location and position through either a global positioning system (GPS) or some local positioning algorithm. These sensor devices are also known as mote, which integrate the processor, communication module and power supply into a single unit [5].

Sensor nodes of WSNs generally use a small, open-source operating system named TinyOS. It is developed by University of California, Berkeley [4].

### B. Cellular Automata (CA)

A cellular automaton (CA) is a collection of cells on a grid of discrete structure. The grid may be one-dimensional or two-dimensional. Each cell in the grid belongs to one of the finite number of states at a time depending on the states of its neighboring cells and the state of a particular cell changes with time. This state changing process, which follows some rule, continues iteratively for desired time steps. The state of the cell may be on/ off or white/ black. The binary, nearest-neighbor, one-dimensional cellular automata are known as "elementary cellular automata". S. Wolfram studied this type of CA. For a three-cell grid there are $2^3 = 8$ possible binary states and a total of $2^8 = 256$ elementary cellular automata. A 3-neighborhood CA can be expressed as [4]

$$\mathbf{x_i(t+1) = f(x_{i-1}(t), x_i(t), x_{i+1}(t))} \qquad (1)$$

where $x_i$ represents a particular cell, $x_{i-1}$ represents left neighbor and $x_{i+1}$ represents right neighbor of $x_i$ respectively, t represents the current time and t+1 represents the next time period, f is the function which implements the CA rule.

Cellular automata can be used in cryptography especially in the public key cryptography domain. The speciality of the finite CA is one way function whose inverse is very hard to find. For a given rule, the future states can be found easily but the reverse is very hard to get. However, one can design the rule in such a manner that it can be inverted easily.

### C. Reversible Cellular Automata (RCA)

A reversible cellular automaton is a special type of CA where every current configuration has exactly one past configuration. In RCA the new state of a cell is determined not only by the present states of the cell and its neighbors but also by the previous state of that cell. A 3-neighborhood RCA can be expressed as [4]

$$\mathbf{x_i(t+1) = f(x_{i-1}(t), x_i(t), x_{i+1}(t), x_i(t-1))} \qquad (2)$$

where $x_i(t-1)$ represents the previous state of the cell $x_i$. The reversible cellular automata are not the invert rules of the original ones.

## III. RELATED WORK

### A. Key Management

The essential requirement for secure data communication is a pair of secret keys. After the initial deployment of a sensor node, the secret key is required to communicate with the neighboring nodes. This phase where the secret key is established for a newly deployed sensor node is known as "Key pre-distribution". But the sensor resource constraints like limited power, limited computation ability or low memory make this process very difficult.

There are several key establishment schemes proposed by several researchers. The pre-distribution scheme can be deterministic or probabilistic. Apart from that there are In-situ key establishment schemes also. In the key pre-distribution schemes, before the deployment, each sensor node contains the keys or keying information. As the network topology is very much unpredictable to any sensor node before deployment, the extra information regarding key is required here. This uncertainty may hamper the performance of the key pre-distribution schemes [6]. Eschenauer and Gligor [2] proposed that every sensor node will have random pre-distribute keys loaded in it before the deployment. These keys will be selected from a large pool of symmetric keys. At the time of communication, the two sensor nodes search that whether they have any common key or not and if found, they share the key. If common key is not found, then they establish a path key. But in this scheme, the attackers may gather the idea and prepare the total key pool after considering several sensor nodes.

Chan et al [7] proposed a scheme where two nodes require q > 1 number of common keys for establishing a shared key. The problem in this scheme is that as more than one pair of nodes may know the keys, so the key may be revealed by capturing a node.

Besides these, two random key space schemes have been proposed by Du et al [8] and Liu and Ning [9]. First one uses symmetric matrices to define the key space and second one uses symmetric polynomials. The sensors, which want to communicate, compute the shared key if they have keying information from the same key space. But, modular multiplications make the schemes expensive in terms of computation, which is a constraint in sensor node. Another random key scheme is proposed by Chan and Perrig whose name is PIKE [10]. Here two sensor nodes communicate with each other with unique pair of keys and if no shared key is present then a trusted intermediary establishes the key.

Another type of key management scheme is in-situ key establishment scheme. In this scheme, sensors compute shared keys with their neighbors after deployment [6]. But like the random key schemes of [8] and [9], in this scheme also, the shared key computation process is complex. Another scheme proposed by Perrig et al is SPINS [3], which uses a trusted base station to establish keys among the sensor nodes.

Teymorian et al. [6] proposed a cellular automaton based key management scheme, CAB, which supports rekeying activities. In CAB, each sensor contains a small number of CAs before deployment. Using CAs, highly random pair-wise keys can be generated using simple bitwise OR and XOR operation. In addition to that, the keys can be computed on demand by sensors that have a common CA.

### B. Secured Data Communication

The method of secure data transmission in WSN is more difficult than other conventional wireless networks because of its constraints mentioned above. There is no adequate memory space where the variables of the asymmetric cryptosystem will be stored. As far as different type of attacks in WSN are concerned, the expected attacks in conventional wireless network like the Denial of service (DoS) attacks, time synchronization attacks, physical layer attacks, routing threats, malicious traffic injection etc. can be possible in WSN also [4].

Roy Chowdhury et al [4] proposed a cellular automaton based lightweight data authentication protocol (CADA) which not only ensures secure data transmission but also tracks malicious motes. In their protocol, the maximum workload is given to the sink keeping in the mind the limited computational capabilities of the motes.

## IV. CAKD - KEY PRE-DISTRIBUTION ALGORITHM

While developing the algorithms, we have kept in our mind the constraints of the sensor nodes that is

a) Nodes have computational constraints i.e. complex computation should be avoided

b) Nodes have small amount of memory, so huge data can not be stored in the nodes

c) Nodes get their energy/ power from battery, so extra computation will decrease the battery power as well as node power also.

Keeping these constraints in mind we have tried to develop the algorithms, which have less complex calculation, take little time also. We have developed the algorithms using reversible cellular automata (RCA) and used one special property of RCA that is [4]

$$\mathbf{RCA\big(X,\,RCA(X,\,Y)\big)=Y} \qquad (3)$$

where X and Y represent previous and current configuration respectively [4].

In a WSN, there may be several clusters of nodes. Each cluster contains a cluster head or base station which acts as server and all the motes in that cluster act as clients of that base station. If one mote wants to communicate with another mote within the same cluster, then the communication is done through the base station. If one mote wants to communicate with a mote of a different cluster, then the communication is done through the base stations of the two clusters. The entire key pre-distribution operation is depicted in figure 1.

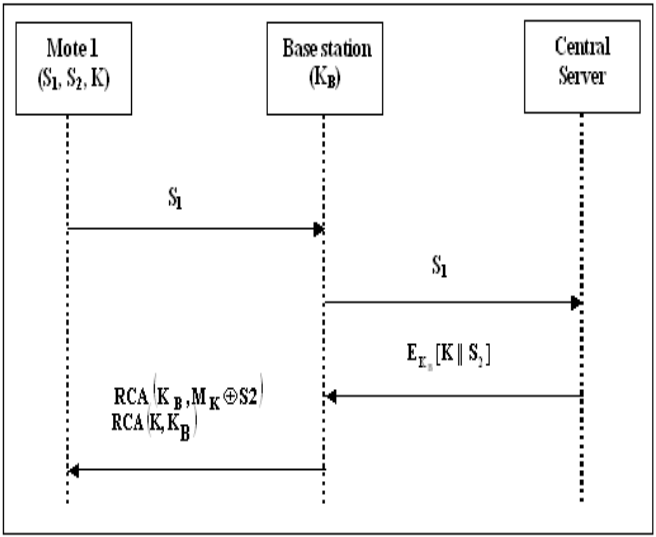

Fig. 1. Distribution of operating key of a sensor node by the base station

a) At the time of deployment, a new mote will contain an encrypted key (K) of 4 bytes and two unique serial numbers S1 and S2 of 3 bytes each. Every mote will have these preloaded information which will be kept by a central server. After the deployment, a new mote M1 sends the serial number S1 to its base station.

b) The base station sends the serial number S1 to the central server for the authentication. If any rouge mote sends a fake serial number, the central server can identify this. If the number is valid then central server will return an acknowledgement to the base station along with preloaded key K of mote M1 and the second serial number S2 encrypted using the base station's key $K_B$ of 8 byte size which is denoted by P in (4). E represents the encryption process. In case of any rouge mote, central server will send the failure message to the base station.

$$\mathbf{P = E_{K_B}\big[K \parallel S2\big]} \qquad (4)$$

c) The base station after getting the appropriate acknowledgement from the central server will first decrypt P to get K and S2 and then will generate a 8 byte mote key $M_K$ and perform an X-OR operation with S2. Then it will perform RCA operations as (5) and (6) and send them to mote M1.

$$\mathbf{A = RCA\big(K_B,\,M_K \oplus S2\big)} \qquad (5)$$

$$\mathbf{RCA\big(K,\,K_B\big)} \qquad (6)$$

d) Mote M1 will perform RCA operations to get the base station key $K_B$ and its own operating key $M_K$ as (7), (8) and (9).

$$\mathbf{K_B = RCA\big(K,\,RCA(K,\,K_B)\big)} \qquad (7)$$

$$\mathbf{M_K \oplus S2 = RCA\big(K_B,\,A\big)} \qquad (8)$$

$$\mathbf{M_K = S2 \oplus \big(M_K \oplus S2\big)} \qquad (9)$$

An adversary needs to know both K and S2 to decrypt the keys $M_K$ and $K_B$. The mote M1 has sent only S1 to the base station after deployment. So if any adversary intercepts the message, he/ she will not get K and S2. Again, the two serial numbers and the key will be used only once because after getting the operating key $M_K$, mote will use that key only for data transmission, which will be described, in the next section.

## V. CASC - SECURED DATA COMMUNICATION ALGORITHM

After the key pre-distribution phase, every cluster's base station will contain the pre-distributed operating keys of all the motes. In addition to that, every base station will contain the secret keys of other cluster's base stations also. On the other hand, all the motes within the cluster contain only a pair of keys, viz. the key of their base station and their own secret encrypted operating keys.

a) Let one mote $M_1$ wants to communicate with mote $M_2$. $M_1$ will generate a random key R of 8 bytes that will work for that session only. The total thing has been depicted in figure 2. After generating the key, $M_1$ will encrypt the data by some encrypting function E and encrypt the random key R using its own pre-distributed key $M_{K1}$ and base station's key $K_B$ as (10) and (11) and then send the encrypted data and encrypted key $R''$ to base station $M_B$:

$$\mathbf{R'_1 = RCA\big(R,\,M_{K1}\big)} \qquad (10)$$

$$\mathbf{R''_1 = RCA\big(K_B,\,R'_1\big)} \qquad (11)$$

b) After receiving the packets from $M_1$, base station will only decrypt the key R as (12) and (13):

$$R'_1 = RCA\left(K_B, R''_1\right) \quad (12)$$

$$R = RCA\left(M_{K1}, R'_1\right) \quad (13)$$

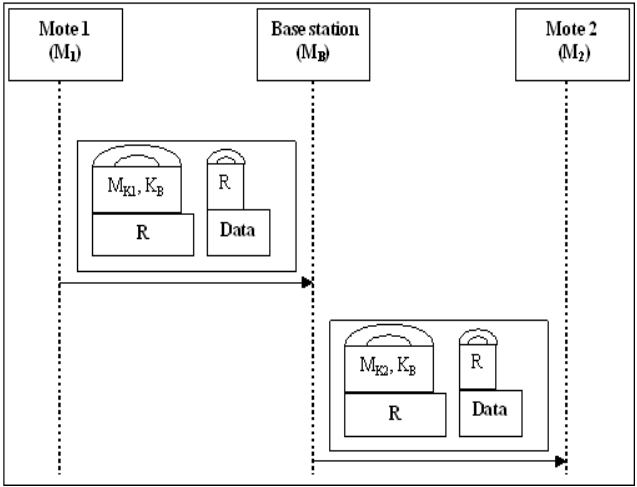

Fig. 2. Data communication between two motes using random key R

Then it re-encrypt the random key R using its own key and destination motes i.e. M2's key $M_{K2}$ as (14) and (15):

$$R'_2 = RCA\left(R, M_{K2}\right) \quad (14)$$

$$R''_2 = RCA\left(K_B, R'_2\right) \quad (15)$$

After that the cluster head will send the newly encrypted key R (in the form of $R_2''$) and the encrypted data to mote $M_2$.

c) $M_2$ will first decrypt the random key R as (16) and (17) and then will decrypt the original message using R:

$$R'_2 = RCA\left(K_B, R''_2\right) \quad (16)$$

$$R = RCA\left(M_{K2}, R'_2\right) \quad (17)$$

d) Apart from that, if any mote wants to communicate with a mote of another cluster then first it will generate the random key, encrypt it using (10) and (11) and send it to its base station along with the destination address. The base station will find the location of the destination, then encrypt the key again using its own and destination base station's keys and will send it to destination base station. Destination base station will work accordingly and send the encrypted data and session key to destination mote using the above formulas.

e) Beside this, every base station will periodically refresh the keys of the motes and send them by encrypting it as (18) and (19):

$$M'_{K1} = RCA\left(M_{K1}^{New}, K_B\right) \quad (18)$$

$$M''_{K1} = RCA\left(M_{K1}, M'_{K1}\right) \quad (19)$$

The mote will decrypt it as (20) and (21), discard the previous key and store this new key. The data communication will be done using this new key until the base station will generate the next key and send that to the mote

$$M'_{K1} = RCA\left(M_{K1}, M''_{K1}\right) \quad (20)$$

$$M^{New}_{K1} = RCA\left(K_B, M'_{K1}\right) \quad (21)$$

where $M_{K1}$ is the previous key of mote $M_1$ and $M_{K1}^{New}$ is the newly generated key by the cluster head.

## VI. ANALYSIS

### A. CAKD algorithm

Most of the key pre-distribution algorithms either uses complex computation or takes much memory space of the motes where both of these are constraints of sensor nodes. According to our algorithm, each sensor node will have to store total 26 bytes of data, one 4 byte key and two 3 byte serial numbers which will be pre loaded in the mote before deployment, one 8 byte pre-distributed key for the mote itself and one 8 byte base station's key. In the contrast, another CA based algorithm CAB [6] needs at least 150 bytes of storage to acquire 90% key sharing probability. Figure 3 shows the comparison between CAB and CAKD with respect to space requirement. Again, there is only one time requirement for the pre-loaded key and two serial numbers, so if any adversary captures any mote, this information will not help him to find out the present keys of the mote. In case of pre-distributed key of the mote and the base station's key, as they are 8 bytes in size, so $2^{64}$ combinations are possible for each key, which is a huge number. Not only that, the base station will also refresh these two keys periodically, so after a period of time, the adversary will find a new pair of keys. In case of mote failure, after the recovery, the mote will start form the beginning using CAKD algorithm with pre-loaded K, S1 and S2.

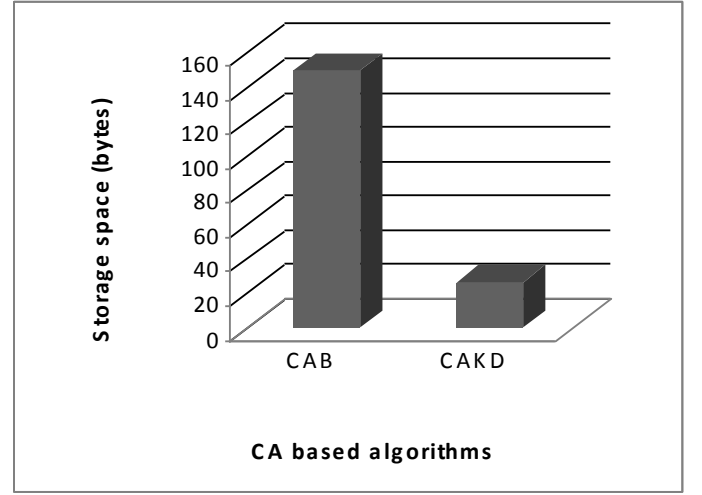

Fig. 3. Comparison between CAB and CAKD with respect to memory space requirement

The operations of key encryption and decryption are also easier because here we have used cellular automata based computations which uses AND, OR, XOR like simple bit-wise operations.

### B. CASC Algorithm

The secured data communication algorithm uses one 8 byte randomly generated session key for each mote. So $2^{64}$ different combinations are possible for each session key. And that key will be used for a single session only. Beside this, this algorithm has the following properties:

a) As the session key is generated randomly and is encrypted by two pre-distributed keys viz. the mote key and the base station's key so if any adversary somehow manages to intercept the packet from the path, he/ she will not be able to decrypt the key because for that he/ she has to know both the pre-distributed keys which is nearly impossible.

b) As the pre-distributed keys of each mote will be periodically changed by the base station, that will prevent **replay attack**, because if the same message is sent later by the adversary, at that time there will be a high probability that key will be changed which will make the decryption process impossible.

There is another way to prevent replay attack that is the source mote can attach a time-stamp with its data that will be checked and verified by the base station and the destination mote.

c) This algorithm can prevent **Man-in-the-middle attack** also. If any intruder wants to enter between the sender and the receiver, he/ she must know all the pre-distributed keys of motes and base station which is nearly impossible, because to send a fraud message to the receiver after intercepting the original message from the sender, the intruder needs to generate another random session key and needs to encrypt that using both the sender's and receiver's key. If he/ she uses his/ her own key, that will not be known by neither the sender nor the receiver.

d) If any unauthorized mote wants to communicate with the base station or to any other authorized motes, this can be prevented also using this algorithm, because that unauthorized mote has to encrypt its session key using its own key and the base station's key. If it somehow manages to get the base station's key then also communication cannot be done because the base station will not have its entry in its own database.

e) As the scheme is very much dependent upon the base station like SPINS [3], so the base station failure may be a big problem. In that case, a provision of a back-up base station will be there. A mote with higher configuration than the other mote can act as back-up base station. In case of main base station failure, that mote will act as the base station automatically for the time being.

### VII. CONCLUSION

In this paper, we have proposed CAWS, a set of algorithms for key pre-distribution and secured data communication of WSNs viz. CAKD and CASC respectively using reversible cellular automata. We have also shown that how the algorithms perform, how they are able to prevent different types of attacks and how they handle different constraints of WSNs efficiently. In the future, we will try to improve CAWS more in terms of lesser utilization of memory and lesser computation complexity.

### REFERENCES

[1] Indrajit Banerjee, Sukanta Das, Hafizur Rahaman and Biplab K Sikdar , "C A Based Sensor Node Management Scheme: An Energy Efficient Approach"; in International Conference on Wireless Communications, Networking and Mobile Computing, 2007. WiCom 2007, pp: 2795 - 2798.

[2] L. Eschenauer and V. D. Gligor, "A key-management scheme for distributed sensor networks," in CCS '02: Proceedings of the 9th ACM conference on Computer and communications security. ACM Press, 2002, pp. 41–47.

[3] Perrig, A., Szewczyk, R., Wen, V., Culler, D., and Tygar, J.; SPINS: Security protocols for sensor networks. J. Wireless Nets. 8, 5 (Sept. 2002) pp. 521–534.

[4] Atanu Roy Chowdhury, Somanath Tripathy and Sukumar Nandi, "Securing Wireless Sensor Networks against Spurious Injections" in 2nd International Conference on Communication Systems Software and Middleware, 2007. COMSWARE 2007 pp. 1 - 5.

[5] J Hill and D Culler "Mica: A wireless platform for deeply embedded networks", IEEE Micro 22, 6 (Nov/Dec 2002).

[6] Amin Y. Teymorian, Liran Ma, Xiuzhen Cheng, "CAB: A Cellular Automata-Based Key Management Scheme for Wireless Sensor Networks" in the Proceedings of MILCOM'07, IEEE, October 2007.

[7] H. Chan, A. Perrig, and D. Song, "Random key predistribution schemes for sensor networks," in SP '03: Proceedings of the 2003 IEEE Symposium on Security and Privacy, 2003, p. 197.

[8] W. Du, J. Deng, Y. S. Han, and P. K. Varshney, "A pairwise key pre-distribution scheme for wireless sensor networks," in CCS '03: Proceedings of the 10th ACM conference on Computer and communications security. ACM Press, 2003, pp. 42–51.

[9] D. Liu and P. Ning, "Establishing pairwise keys in distributed sensor networks," in CCS '03: Proceedings of the 10th ACM conference on Computer and communications security, 2003, pp. 52–61.

[10] H. Chan and A. Perrig, PIKE: Peer intermediaries for key establishment in sensor networks," in Infocom 2005: The $24^{th}$ Conference of the IEEE Communications Society, 2005.